\documentstyle[12pt,epsfig]{article}
\begin{document}


\thispagestyle{empty}
\renewcommand{\thefootnote}{\fnsymbol{footnote}}

\setcounter{page}{0}
\begin{flushright}
DESY 95--017\\
\end{flushright}

\begin{center}
\vspace*{0.5cm}
{\Large\bf
 NON-SINGLET STRUCTURE FUNCTIONS AT SMALL X
}\\
\vspace{1.4cm}
{\sc  B.~I.~Ermolaev\footnote{This work was supported in part by the
grant R26000 from the International Science Foundation  and in
part by Volkswagen Stiftung},
S.~I.~Manayenkov \footnote{This work was supported in part by the Soros
grant awarded by International Science Foundation},
M.~G.~Ryskin\footnote{This work was supported in part by the grant
INTAS-93-79}}\\
\vspace*{0.3cm}
{\it Petersburg Nuclear Physics Institute,\\
Russian Academy of Science,\\
Gatchina, St.Petersburg district, 188350, Russia\\}
\vspace*{0.3cm}
{\bf Abstract}\\
\end{center}
\parbox[t]{\textwidth}
{The small x behaviour of the non-singlet structure function is
studied within the double logarithmic approximation (DLA) of  perturbative
QCD. Since there is neither $k_T$ nor $\theta$
ordering in
the ladder Feynman graphs,  the predicted non-singlet
quark
densities for the HERA kinematical range ($x\sim 10^{-3}$) exceed the values
calculated from the small-$x$ approximation  of the conventional
Altarelli-Parisi evolution by a factor up to ten.}
\vspace*{0.3cm}
\newpage
\renewcommand{\thefootnote}{\arabic{footnote}}
\setcounter{footnote}{0}

\newpage
\renewcommand{\thefootnote}{\arabic{footnote}}
\setcounter{footnote}{0}

\section{INTRODUCTION}

The small x behaviour of the non-singlet structure functions in  deep
inelastic scattering plays an important role for the precise description of
the quark densities. Indeed, in order to check the Gottfried sum rule or to
estimate the fraction of the initial nucleon spin carried by the quark (the
spin crisis problem) one has to extrapolate the data into the small x
region,
and therefore one has to know the behaviour of the structure functions at small
$x$.

The kinematical region $x \ll 1$ is of the Regge type, hence there are two
possibilities for its investigation:

(i) to use conventional phenomenological formulae of the Regge type for the
structure function description at the hadron level (see e.g. the review
\cite{RYSK} and references therein);

(ii) within the framework of the perturbative QCD, to calculate and sum up the
Feynman graphs yielding the leading contributions to the structure functions
at the partonic level.

In the latter case one should take into account the Feynman graphs to all
orders
in the coupling constant for $x \ll 1$. The most popular way to carry
this out
consists of extrapolating (see e.g.  \cite{INGEL}, \cite{MARWEB},
\cite{FORBAL})
the Altarelli-Parisi equations, which describe DGLAP evolution of the
structure functions with
respect to $ln(Q^{2})$ \cite{GLIP}, \cite{ALPAR}, \cite{DOK}
into the region $x \ll 1$. This
equation sums up in the region $x \sim O(1)$
all the leading contributions coming from the Feynman graphs of the ladder
type for the kinematical region where the transverse momenta $k_{jT}$
of virtual partons (quarks and gluons) are strongly ordered
\begin{equation}
k_{1T} \ll k_{2T} \; \ll\;...\;\ll k_{nT} \ll \sqrt{Q^{2}}.
\end{equation}
The quantity $Q^{2}=-q^{2}>0$ denotes the   square of the
photon mass.

Indeed, it has been proven in \cite{DOK} that an integration over
$k_{j}$ in the
region (1) yields the main contribution to the singlet and non-singlet
structure functions at $x \sim O(1)$.

However, such ordering is not valid for summing up the leading contributions
when $x \ll 1$. In this case the well
known result  \footnote{Firstly, this result has been obtained in
\cite{GGLF} for the investigation
of the process $e^{+}e^{-} \rightarrow \mu ^{+} \mu ^{-}$ in QED. See also the
review \cite{GOR}.} is that the condition (1) should be replaced by the
ordering of the
longitudinal Sudakov variables $\alpha _{i},\beta _{i}$ for momenta $k_{i}$
as follows:
\begin{eqnarray}
\nonumber
\alpha _{1} \ll \;\alpha _{2} \ll \;...\;\ll \alpha_{n} \ll 1,\\
1 \gg \beta _{1} \gg \;...\; \gg \beta _{n}.
\end{eqnarray}
Integration over $k_{i}$ in the region (2) yields the leading contributions
to the structure functions, in particular, the double-logarithmic (DL)
contributions for the non-singlet structure functions (NSSF). Thus, one can
evaluate NSSF by the direct calculation of the Feynman graphs of the ladder
type. However there is another, simpler way to do this, namely, to make use of
the Infrared Evolution Equation (IREE) for NSSF, i.e. of the evolution
equation with
respect to the infrared cut-off in the transverse momentum space. This method
has first been applied in \cite{KLIP} to the description of the elastic
scattering of quarks, and it has then been
generalized to the investigation of inelastic scattering in QED and QCD (see
  \cite{KLIP}, \cite{ELIP} ,\cite{ERM}  and references therein).On the
whole, this method has turned out to be an efficient way to calculate
 scattering amplitudes at high energies.

In the present paper we calculate the non-singlet structure functions  in
the double logarithmic approximation (DLA) by constructing
and solving both the integral equation of the Bethe-Salpeter type and IREE
for it. In contrast to
calculations based on extrapolating  the conventional DGLAP  into the
region $x \ll 1$, we take into account both the contributions $\sim
[\alpha _{s}\;ln(x)\;ln(Q^{2})]^{n}$ and the DL-contributions $\sim (\alpha
_{s}
ln^{2}x)^{n}$ in every order in  $\alpha _{s}$.
The latter ones  become even more important at very small $x$.

The paper is organized as follows.
In Section 2 we review the main features
of calculating  Feynman graphs in DLA. We derive the linear integral
equation for NSSF, taking into account the running QCD coupling constant
$\alpha _{s}(Q^{2})$ within DLA.
In Section 3 we construct and solve IREE for
NSSF. Here we assume $\alpha _{s}$ to be fixed.
The solution is given in terms of the Mellin transformation.
In Section 4 we discuss our results and compare them with the
conventional expression for NSSF obtained with Altarelli-Parisi evolution
equation.

\section{INTEGRAL EQUATION FOR THE NON-SINGLET STRUCTURE FUNCTION}

In this section we give the outline of the Feynman graph calculation in
DLA developed firstly in \cite{GGLF} ( see also the review \cite{GOR})
for the forward
$e^+e^- \rightarrow \mu ^+ \mu ^-$ scattering in QED and use it for
constructing the integral equation for the non-singlet structure function
$f _{NS}(x, Q^2)\;(f_{NS} \equiv q_{n.s.}(x,Q^2))$ in  deep inelastic
lepton-nucleon scattering (DIS) at small x. In the framework of the
perturbative QCD $f_{NS}$ is given by the graph in Fig. 1 where the dashed
lines
correspond to the virtual photon with momentum q  and
the straight lines correspond to the quarks. The wavy lines we assign for
gluons. The blob in Fig. 1  means that the radiative corrections in DLA are
taken into account. The dotted line crossing the blob means that the
intermediate s-channel particles are on-shell. We  investigate
$f_{NS}$ in the kinematical region
\begin{equation}
\mu ^2 \ll Q^2 \ll 2pq
\end{equation}
i.e. at small x, where $x=Q^2/2pq$, $p$ denotes the initial quark
momentum and $\mu$ is the infrared cut-off.
The leading (the double-logarithmic) contributions to $f _{NS}$
come from the ladder Feynman graphs (Fig. 2). Non-ladder graphs, though
yield DL-contributions, do not contribute to $f _{NS}$ because they
compensate
each other totally. The reason for such a compensation in QCD and QED is
that all non-ladder gluons are nearly on-shell and can be treated like the
bremsstrahlung in the forward scattering, cancelling each other
due to the destructive interference.

Thus, DL-contributions to the non-singlet structure function come from the
ladder Feynman graphs in Fig. 2 only. To evaluate them, one should solve
the linear integral
equation in Fig. 3. for the sum of DL-corrections, A, to the elastic forward
quark-antiquark scattering amplitude $A(t, \beta)$ (see Fig. 3):
\begin{eqnarray}
A(t, \beta) = \int \limits _{\beta} ^{1} \frac {d \beta ^{'}}{\beta {'}}
\int \limits _{ \mu^2}^{t_m} \frac{dt^{'}}{t^{'}}A(t^{'} ,\beta {'})\frac
{C_F}{2 \pi} \alpha _s(t_m)+A_0,
\end{eqnarray}
where $\alpha_s$ is the QCD-coupling constant, $C_F=(N^2-1)/2N$ for
the
colour group SU(N). We have used in (4) the Sudakov parameterization for $k$
in the following form
\begin{equation}
k= \alpha q'+ \beta p+k_T,\;\; -k_T^2 \equiv t > 0,
\end{equation}
where
\begin{equation}
 q'_{\mu}=q_{\mu}- \frac{q^2 p_{\mu}}{2pq}\;,
\end{equation}
and the analogous parameterization for $k'$.
The nonhomogeneous term, $A_0$, in (4) is
\begin{equation}
A_0(t, \beta)= 4 \pi C_{F} \alpha _{s}(t/ \beta).
\end{equation}
The limits in the integral over $\beta ^{'}$ in (4) corresponds to the
$\beta$-ordering (see e.g. \cite{GOR});
\begin{equation}
 \beta  ^{'} > \beta \;.
\end{equation}
The lower limit in the integral over $t^{'}$ is given by some infrared
cut-off
$\mu$, originated by applicability of the perturbative QCD to the problem
, while the upper one comes from the condition $k^2_{\parallel}\equiv
\alpha \beta s \leq \mid t \mid$ (where $s \equiv 2pq$). Indeed, if $k^{'}_T
\gg k_T$ the value of $\alpha$ in (5) is $\alpha \approx (k_T-k^{'}_T)^2/
\beta ^{'}s \approx t^{'}/s \beta^{'}$ and, when $t^{'} \gg t \beta^{'}/\beta$,
the longitudinal part of $k^2_{\parallel}$ becomes so large,
\begin{equation}
k^2_{\parallel} =s \alpha \beta \sim t^{'} \beta/\beta^{'} \geq t,
\end{equation}

that it
destroys the logarithmic structure of the integrand $dt^{'}/t^{'}$ in (4)
(see \cite{GOR} for the details).

So, within the DLA we put
\begin{equation}
t_m=t \beta^{'}/ \beta\;.
\end{equation}
Let us notice that for small x (i.e. when $\beta^{'} \gg \beta$) this
value
is  larger than the upper limit when $t_i$ are ordered.
Now let us show that the argument of the QCD coupling constant in (4) is
also equal to $t_m$. To find it, one has to consider the quark loop
inserted
into the gluon propagators in Fig. 3 (see also Fig. 4). The integral over
the invariant mass of the loop looks as $\int dM^2/M^2$. This logarithmical
behaviour continues up to $M^2 \sim t_m=t \beta^{'}/ \beta$ where the large
value of $M^2$ corresponds again to the essentially large $k^2_{\parallel}
\sim M^2 \beta / \beta^{'} \sim t$. Thus, the argument of $\alpha _s$
should be $t_m$
\footnote{Note that for small x (i.e. when $\beta^{'} \gg \beta$) this value
is much greater than the argument in the canonical Altarelli-Parisi equation
($t_m=t$).}.

Finally, to obtain the non-singlet structure function, $f_{NS}$, we integrate
$A$ over $t$:
\begin{equation}
f_{NS}(x ,Q^2)=e^2_q \Bigl\{\delta  (1-x) +\frac{1}{8 \pi ^2} \int\limits
^{s _{m}}_{\mu ^2} A(t, \beta ) \frac{dt}{t}\;\Bigr\}\;,
\end{equation}
with $ s_{m} =s/4$ and
 \begin{equation}
\beta =x+t/s\;.
\end{equation}
$e_q$ in (11) denotes the electric charge of the initial quark.
Here the upper limit, $s/4$, is fixed formally by the kinematics, but, in
reality, the bulk of integration in (11) is determined by dependence on
$\beta$. At $t \gg Q^2$, the term $t/s$ increases the value of $\beta$
crucially, and as the amplitude $A(t, \beta)$ falls down with an increase
of $\beta$ the essential value of $t$ in equation (11) is of the order of
$Q^2$.
It should be noticed that the expression (11) corresponds to the integration
with the transversely polarized photon i.e. to $\sigma^T$. In the case of
the longitudinal heavy photons one has to write, in the denominator,
$Q^2$ instead of $t$
\begin{equation}
\frac{\sigma^L}{\sigma^T}=\frac{\int\limits^{s_{m}}_{\mu^2}A(t,\beta )
dt/t}
{\int\limits^{s_{m}}_{\mu^2}A(t,\beta )dt/Q^2 },
\end{equation}
where  $\beta $ in (13) is given by (12).

\section{IREE FOR THE NON-SINGLET STRUCTURE FUNCTION}
 In this section we construct and solve the evolution equation for NSSF
 with respect to the infrared cut-off in the transverse momentum space.
 It is more convenient to consider, instead of the non-singlet structure
 function, $f_{NS}$, the scattering amplitude, $M$, for the forward
 quark-photon  scattering where photon is off-shell (see Fig.4). The
 Optical theorem gives the simple relation between
 $f_{NS}$ and $M$.

 Indeed,
\begin{equation}
4 \pi ^2 \alpha \frac{tr\{\hat p \hat e (\hat p+ \hat
q) \hat e \}}{(2pq)^2} f_{NS}=\frac{1}{s} Im M
\end{equation}
where   $e_{\mu}$ is the photon polarization vector and $\alpha$ is
the QED coupling constant.
As we are discussing the forward scattering, the amplitude $M$ depends on
 $s,\;Q^2$ and on the masses of quarks taken into account.

 There are  graphs with the ultraviolet and the infrared
 singularities among the Feynman graphs contributing to $M$. We can easily
avoid the problem of the
 ultraviolet singularities by noticing that we will treat any single Feynman
 graph contribution to $M$ as the result  obtained with the dispersion
 relations from its imaginary part (with respect to s) which has not the
 ultraviolet singularities. In order to get rid of the infrared
singularity
 problem, one should introduce some infrared regularization. Let us define
 the infrared cut-off parameter, $\mu$:
\begin{equation}
\mu \ll k_{iT},\;\;\;i=1,\;2,...
\end{equation}
where $k_{iT}$ denotes the momentum of virtual particle i (a quark or
gluon), transverse to the plane formed by momenta $p$ and $q'$ (see (5)):
\begin{equation}
k_{iT}p=k_{iT}q^{'}=0.
\end{equation}
Such a regularization does not destroy the gauge invariance, so we can fix
any gauge for the virtual gluons at convenience. In the present paper we
 use the Feynman gauge.

It was shown in \cite{ELIP} that $\mu$,  being the minimal transverse
momentum,
also takes part of a new mass scale. Thus, we can neglect masses of the
virtual
quarks and to be free of the infrared singularities in the same time. In
DLA, the value of $\mu$ can be chosen much greater than $\Lambda _{QCD}$.

As $M$ now depends on $s,\;Q^2$, and on the new mass scale $\mu$, in the
form \begin{equation}
M=M(s/ \mu ^2, Q^2/ \mu ^2),
\end{equation}
one can differentiate it with respect to $\mu$. Let us notice
that
\begin{equation}
-\mu^2\frac{\partial M}{\partial \mu ^2}=\frac {\partial M}{\partial z}+
\frac{\partial M}{\partial y}
\end{equation}
where
\begin{equation}
z=ln(s/ \mu^2),\;\;y=ln(Q^2/ \mu^2).
\end{equation}
Expression (18) is the left-hand side of the IREE for $M$ (the left-hand
side of the equation in Fig.5).

Now let us obtain the right-hand side of the IREE. In the first place,
let us notice that the ordering conditions (8) do not mean the ordering
for $t_i$ in the ladder Feynman graphs. Indeed, integration of such graphs
over $\alpha_i$ yields (see \cite{GOR})
\begin{equation}
t_1/ \beta _1 \ll t_2/ \beta _2 \ll \cdots \ll t_n/\beta _n.
\end{equation}
Since $t_i$ are not ordered, integration over any of $t_i$ gets $\mu ^2$
as the lowest limit. However, the region (20) can be considered  as a
superposition of the regions $\Omega _i$ where the transverse momenta
are ordered, with $t_i\;\; (i=1, \cdots ,n$) being the minimal transverse
momenta in the region $\Omega _i$. Therefore, in every such region
$\Omega _i$, $\mu$ is the lowest limit only for integration over $k_{iT}$
whereas integration over other transverse momenta does not involve $\mu$
at all. In other words, the ladder Feynman graph ( and therefore the sum of
them) can be treated as a convolution of two ladder graphs, so that, in each
the graph, integration over  transverse momenta begins from
$t_i$ in the region $\Omega _i$, i.e. for them $k_{iT}$ takes place of
$\mu$, being a new infrared cut-off when the quark masses are neglected.
Let us denote such a minimal transverse momentum as $k_T$. Ordering (20)
means that the quark pair with the minimal $k_T$ can be in any place of
the ladder graph in Fig. 2. In the case of $t_i$-ordering the quarks with
the minimal $k_T$ would be in the fixed place of the ladder graph (in the
bottom of the graph in Fig. 2).

Thus, beyond
the Born approximation, the amplitude $M$ can be represented as the
convolution of two amplitudes (the last term in the right-hand side of
the equation in Fig. 5). The first of these amplitudes is the same
amplitude $M$ with replacement $\mu$ by $k_T$ and $s$ by $2kq$. The
second one is the amplitude $M_0$ for the forward quark-antiquark
scattering.

Adding the Born amplitude $M_B$ to the right-hand side of the  equation,
we obtain  the equation in Fig. 5 for $f_{NS}$ in the graphic form.
The letters inside the blobs in Fig. 5 denotes the infrared cut-offs.
This equation can be easily rewritten in the analytic form in the
momentum space with using the standard Feynman rules.

It is convenient however, to perform the Sommerfeld-Watson
transformation which partly coincides with the Mellin
transformation :
\begin{equation}
M(z,y)=\int\limits ^{\lambda +i \infty}_{\lambda -i \infty} \frac{d
\omega}{2 \pi i}(s/ \mu ^2) ^{\omega} F(\omega ,y)
\end{equation}
where $\lambda$ is chosen so that the integtation contour in (21) was on the
right of the singularities of $F$.
The variable $\omega$ corresponds to the angular momentum in the complex
momentum plane. Having performed the Mellin transformation, we obtain the
equation for $F$ in the following simple form:
\footnote{For details of the method see \cite{KLIP}, \cite{ELIP},
\cite{ERM} and the references therein.}
\begin{equation}
\omega F+\frac{\partial F}{\partial y}=\frac{1}{8 \pi ^2} f_0(\omega) F
\end{equation}
where $f _0$ is the Mellin amplitude for the forward scattering of on-shell
quarks in DLA \cite{KLIP}. There is not a contribution coming from the
Born
amplitude $M_{B}$ in (22) because $M_B$ does not depend on $\mu$ in the
region (2).

As equation (22) is linear, one can easily obtain its solution:
\begin{equation}
F=\Phi(\omega) \exp \{ (-\omega +f_0/8 \pi ^2)y \}
\end{equation}
$\Phi(\omega)$ in (23) should be specified with a certain boundary
condition. Let us fix it at the point $y=0$:
\begin{equation}
F(\omega ,y) \mid _{y=0}=\Phi(\omega).
\end{equation}
Thus, in order to specify $\Phi$ we should construct and solve IREE for the
amplitude of quark-photon forward scattering, $\tilde M$, where the
photon is (nearly) on-shell. Immediately we obtain for its Mellin
amplitude, $\tilde F(\omega)$, similar equation:
\begin{equation}
\omega \tilde F= C_0+\frac{1}{8 \pi ^2} f_0(\omega)\tilde F(\omega)
\end{equation}
\begin{equation}
C_0=\frac{-4 \pi tr\{\hat p \hat e (\hat p+ \hat q ) \hat e\} e^2_q}
{2pq}\;.
\end{equation}
The first term in the right-hand side of (25) is the contribution of the
Born graph which depends on $\mu$ when $y=0$.

Solution of (25) gives us
\begin{equation}
\tilde F=C_{0}/(\omega-f_{0}/8 \pi ^2).
\end{equation}

The Mellin amplitude $f_{0}(\omega)$, firstly obtained
in \cite{KLIP}, satisfies the same
differential equation  like $\tilde F(\omega)$ with obvious replacements
$\tilde F$ by $f_{0}$ in equation (25) and $C_0$ by $a_0$,
\begin{equation}
a_0=g^2C_F
\end{equation}
where $g$ is the QCD-coupling constant.
 Thus, the equation
\begin{equation}
\omega f_0 =a_0+f^2_0/(8\pi^2)
\end{equation}
for $f_0$ is non-linear, though very simple. Its solution is
\begin{equation}
f_0=4 \pi^2 \omega [1- \sqrt{1-a_0/(2\pi^2 \omega ^2)}]
\end{equation}
where the sign minus before the square root sign was chosen so that
\begin{equation}
f_0 \mid _{\omega \gg 1} \sim a_0/ \omega
\end{equation}
where $a_0/\omega$ is  the Born approximation for $f_0$. Eq. (29)
permits to simplify (27) because it gives
\begin{equation}
\omega - f_0/(8 \pi^2) =a_0/f_0
\end{equation}
Thus, combining (29), (27) and (23), we obtain
\begin{equation}
F(\omega ,y)=\frac{C_0}{g^2C_F}f_0(\omega) \exp \{-[\omega-f_0/(8
\pi^2)]y \}.
\end{equation}
It leads to
\begin{equation}
M(\frac{s}{\mu^2}, \frac{Q^2}{\mu^2})=\frac{C_0}{g^2C_F}
\int\limits_{\lambda - i \infty}^{\lambda +i \infty}\frac{d \omega}{2 \pi i}
\Bigl(\frac{s}{Q^2}\Bigr)^{\omega}f_0(\omega) \exp \{yf_0(\omega)/(8 \pi^
2)\}. \end{equation}
With using (14) and the fact that in DLA
\begin{equation}
Im M \approx - \pi dM/dln(s)
\end{equation}
we obtain for the non-singlet structure function the following expression
\begin{equation}
f_{NS}=\frac{  e_q^2}{g^2 C_F}
\int \limits_{\lambda - i \infty}^{\lambda +i \infty}\frac{d \omega}{2 \pi i}
\Bigl(\frac{s}{Q^2}\Bigr)^{\omega } \omega f_0(\omega ) (Q^2/ \mu ^2) ^
{f_0(\omega )/(8 \pi ^2)}.
 \end{equation}

\section{DISCUSSION}
 The expression (36) for the non-singlet structure function takes into
account DL contributions to all orders in the coupling constant, and
therefore can be represented as an expansion in powers of $\alpha _s$.
In terms of the Mellin  variable $\omega$, we can consider the expansion
of $f_0$ into a power series in $1/ \omega$ and integrate the result over
$\omega$ according to
\begin{equation}
\int \frac{d \omega}{2 \pi i}(\frac{1}{x})^{\omega} \omega ^{-k-1}=
\frac{1}{k!}ln^k(\frac{1}{x}),
\end{equation}
$k=1,\;2, \cdots$.

As
\begin{equation}
f_0 \approx 4 \pi ^2 (\frac{a}{2 \omega}+\frac{a^2}{8 \omega
^3}+\frac{a^3}{16 \omega ^5})+O( \omega ^{-7}),
\end{equation}
with $a=2 \alpha _s C_F/ \pi$, we obtain from (36)
\begin{eqnarray}
f _{NS} \approx \frac{4 \pi ^2 e^2_q}{g^2C_F} \int \frac{d \omega}{2 \pi
i}\Bigl(\frac{1}{x}\Bigr)^{\omega} \omega \Bigl(\frac{a}{2
\omega}+\frac{a^2}{8 \omega ^3}+\frac{a^3}{16 \omega ^5}\Bigr)
\nonumber\\ \exp\Bigl\{y\bigl[
\frac{a}{4 \omega}+\frac{a^2}{16 \omega ^3}+\frac{a^3}{32 \omega
^5}\bigr]\Bigr\}\;.
\end{eqnarray}
It should be stressed that for the non-singlet function at small $x$ one
deals with the expansion $(\alpha_s/\omega^2)^k$, which is much more
singular ( at $\omega \rightarrow 0$) than the expansion
$(\alpha _s/\omega )^k$ typical for the singlet (BFKL-pomeron) case.
Thus, the summation of the whole DL series becomes very important.

If
$\omega \gg a$, one can neglect all terms in (39) proportional to $(a/
\omega ^2)^k$ ($k>1$), compared to the first one. It leads to the well known
expression  $\tilde f_{NS}$ for the non-singlet structure function:
\begin{equation}
 f_{NS} \approx \tilde f _{NS} = e^{2}_{q} \int \limits_{\lambda -i\infty}
^{\lambda +i\infty} \frac{d \omega}{2 \pi i}
\Bigl(\frac{1}{x}\Bigr)^{\omega} exp(\frac {ay}{4 \omega}).
\end{equation}
Expression (40) is obtained from the conventional Altarelli-Parisi evolution
equation which is valid in the region $x \sim 1$
where it correctly sums up all the leading $lnQ^2$ contributions to
$f_{NS}$, by retaining in the anomalous dimension only the singular part
in $1/ \omega$. However, as it is clear from (39), it
fails in the region $x \ll1$, where DL-contributions, proportional to
$ln^{2k}(1/x)$ becomes extremely essential and therefore
should not be neglected.

To demonstrate this let us expand (36) and (40) in inverse powers of
$\omega$ and
integrate them over $\omega$ with using (37). We obtain \begin{eqnarray}
\frac{f_{NS}}{e^2_q}= \delta (x-1)+\frac{a}{4}[ln(1/x)+ ln(Q^2/ \mu
^2)]+\nonumber\\+\frac{a^2}{96}[2 ln^3  (1/x)+6 ln^2 (1/x)ln(Q^2/ \mu ^2)+
\nonumber\\+3 ln(1/x) ln^2(Q^2/ \mu ^2)] +O(a^3), \\
\frac{\tilde f_{NS}}{e^2_q}= \delta (x-1)+\frac{a}{4}ln(Q^2/ \mu
^2)+\frac{a^2}{32} ln(1/x)ln^2(Q^2/ \mu ^2) +O(a^3).
\end{eqnarray}
Though the first, the Born terms, in (41), (42) coincide, higher loop
corrections are quite different. Indeed, for the ratio of the first loop
contributions (terms proportional to $a$) we obtain
\begin{equation}
r_1=1+\frac{ln(1/x)}{ln(Q^2/ \mu^2)}\;,
\end{equation}
and for the ratio of the second loop contributions (terms proportional
to $a^2$)
\begin{equation}
r_2=1+\frac{2ln(1/x)}{ln(Q^2/ \mu^2)}+
\frac{2ln^2 (1/x)}{3ln^2(Q^2/ \mu^2)}.
\end{equation}
If we put, for example,
\begin{equation}
\mu ^2=0.1\;GeV^2,\;Q^2=20\;GeV^2\; \mbox{and}\;x=10^{-3},
\end{equation}
we obtain
\begin{equation}
r_1 \approx 2.3,\;\;r_2 \approx 4.7.
\end{equation}
Thus, being equal in the Born approximation, $f_{NS}$ and the
conventional non-singlet structure function $\tilde f_{NS}$ differ a lot
when the radiative corrections are  taken into account. This difference
increases with the order of  perturbation expansion.

Now let us estimate  the ratio $f_{NS}/ \tilde f_{NS}$
 at small $x$, using their
asymptotic expressions. With  the standard technique
(the saddle point method), we find that at $x \ll 1$
\begin{equation}
\tilde f _{NS} \sim \frac{e_q^2}{2 \sqrt{\pi}}\Bigl(\frac {b^2 y}{4 \rho
^3 } \Bigr) ^{1/4} \exp(b \sqrt{y \rho})
\end{equation}
where we have used the notations $\rho =- ln(x);\;b=\sqrt{a},\; y=\ln(Q^2/
\mu ^2)$.

The asymptotic expression for $f_{NS}$ at the same region is
\begin{equation}
f_{NS} \sim \frac{e_q^2 b^2}{2 \sqrt{ \pi}}\Bigl(1-\frac {y}{2
\rho}\Bigr)\sqrt { \frac{b y^3}{4 \rho ^3}} \exp\{b(\rho +y/2-y^2/4 \rho)\}.
\end{equation}
Thus, at $x \rightarrow 0$
\begin{equation}
\tilde  f_{NS}/f_{NS} \sim \frac {1}{2}( \rho /y)^{3/4}\exp\bigl\{b(
\sqrt{ \rho y} - \rho -y/2 +y^2/4 \rho ^2) \bigr \}
\end{equation}
For  example, if we choose the same values (45) for $\mu ^2
,\;s, \;Q^2$  and put  $\alpha _s=0.3$, we obtain from (49)
the estimate
$f_{NS}/ \tilde f_{NS} \approx 7.89$. More accurately, a numerical
evaluation
of $f_{NS}$ and $\tilde f_{NS}$ given by (36) and (40) yields the
following result:
 \begin{equation}
 f _{NS}/\tilde f_{NS} \approx 5.56\;.
\end{equation}
Our results for  other values of $x$ and $\mu$, with $Q^2=20$ $GeV ^2$,
are presented in Table 1. The value of $\alpha _s$ for $\mu ^2 =0.1$ was
put by hands equal to 0.3 whereas for other $\mu$ $\alpha _s = \alpha
_s(\mu ^2)$.
\hspace{0.4cm}
\begin{center}
\begin{tabular}{|l|l|l|l|}
\hline \multicolumn{4}{|c|}{\bf Table 1.} \\ \hline
$\mu ^2, \;GeV ^2$& $x$& $R=f_{NS}/ \tilde f_{NS}$& $\alpha _s$ \\ \hline
0.1&0.001&5.56&$0.3^{\star}$ \\ \hline
0.1&0.01&2.94&$0.3^{\star}$ \\ \hline
0.1&0.1&1.64&$0.3 ^{\star}$\\ \hline
1.0&0.001&10.26&0.394 \\ \hline
1.0&0.01&4.52&0.394 \\ \hline
1.0&0.1 &2.10&0.394 \\ \hline
4.0&0.001&12.71&0.283 \\ \hline
4.0&0.01 &5.94& 0.283 \\ \hline
4.0&0.1  &2.75& 0.283 \\ \hline
\end{tabular}
\end{center}
\hspace{0.4cm}

In the Mellin representation, the Altarelli-Parisi equation for $\tilde
f_{NS}$ can be written as
\begin{equation}
\frac{ \partial \tilde f_{NS}}{\partial ln(Q^2)}=[-\omega + g^2 C_F/\omega ]
\tilde f_{NS}\;.
 \end{equation}
 IREE gives  the $Q^2$-dependence of $f_{NS}$ as:
\begin{equation}
\frac{ \partial f_{NS}}{\partial ln(Q^2)}=[-\omega +  f_0(\omega)]
f_{NS}.
 \end{equation}
The first terms in the brackets in (51), (52) are not essential. They
disappear if one  uses the Mellin transformation in the form
\begin{equation}
M=\int \frac{d \omega }{2 \pi i} \Bigl( \frac{s}{Q^2}\Bigr)^{\omega}
F(\omega, Q^2) \end{equation}
instead of (21).

Thus, the only essential difference between (51) and (52) is the
difference in the "kernels" in these equations. When $x \sim 1$, the
replacement of $f_0$ by its Born value makes (52)  coincide with (51).
In a sense , the opposite replacement $g^2 C_F/ \omega$ by $f_0( \omega )$
corresponds to taking into account all next-to-leading corrections of the
type of $( \alpha _s/ \omega ^2)^n$ to the Altarelli-Parizi kernel. Being
small in the region $ \omega \sim 1$, they become large when $\omega
\rightarrow 0$ ($x \rightarrow 0$). In this connection, let us notice
that these corrections are proportional to $( \alpha _s/ \omega ^2)^k$
only due to the fact that integration over all transverse momenta yields
large contributions from the region of small $k_T\;(k_T \geq \mu)$.
Hence,
if one wants to take into account non-perturbative contributions to
$f_{NS}$ ( see e.g. \cite{FORBAL}), one has to consider them,
besides the leading order, in the non-leading orders also.

 Thus, in
the present paper we have
shown that the applicability of the conventional expression (40) for
the non-singlet structure function at the small x region is quite
doubtful. The standard  and widespread Monte-Carlo programs
(e.g. HERWIG and LEPTO) should be changed when they operate with the $f_{NS}$
in the region $x \ll 1$. The main reason for the descrepancy  is
that in the Regge kinematical region $x \ll1$, the transverse momenta of
the virtual quarks in the ladder graphs are not ordered, in contrast to
the transverse momentum ordering in the Altarelli-Parisi evolution at $x
\sim 1$. It leads to replacement of expression (40) by $f_{NS}$ given by
(36).

Such an amendment leads to a drastic change of the asymptotic
behaviour of the non-singlet structure function. Indeed,  the
conventional formula (40) leads to the behaviour for $x \rightarrow 0$
 \begin{equation}
\tilde f _{NS} \sim \exp\sqrt{ \frac{2 \alpha _s C_F}{ \pi}ln(Q^2/
\mu^2)ln \frac{1}{x}},
\end{equation}
while the IR evolution equation gives at $x \rightarrow 0$
 \begin{equation}
f _{NS} \sim \exp \Bigl\{\Bigl(\sqrt{ \frac{2 \alpha _s C_F}{
\pi}} \Bigr) \Bigl[ \frac{1}{2} ln(\frac{Q^2}{ \mu ^2})+
ln(\frac{1}{x}) \Bigr]\Bigr\}\;,
\end{equation}
thus demonstrating the more singular behaviour of $f_{NS}$ compared to
$\tilde f _{NS}$ at small x. Such a behaviour can be written as
\begin{equation}
f_{NS} \sim x^{- \gamma}.
\end{equation}
where $\gamma$ is an effective power. One can see that, for $\tilde
f_{NS}$, its value\begin{equation}
\tilde \gamma =
\sqrt {\frac{2 \alpha _s C_F}{ \pi}\frac{ln(Q^2/ \mu ^2)}{ln(1/x)}}
\end{equation}
is smaller at very small $x$ than $ \tilde \gamma $ from (55) which predicts
\begin{equation}
\gamma= \sqrt{2 \alpha _s C_F/ \pi}\;.
\end{equation}

\section*{ACKNOWLEGEMENT}
We are grateful to J.Bartels and to L.N.Lipatov for useful discussions. We
would like to gratefully acknowledge the hospitality of
the DESY Theory Group and the financial support of the Volkswagen Stiftung.

\newpage
\section*{Figure captions}
\hspace{3 mm}
Fig.1. Graph for $f_{NS}$. The blob indicates that the radiative corrections
in DLA are taken into account.
\vspace {3mm}
\hspace{3 mm}
\newline
Fig.2.
Graph of the ladder type for $f_{NS}$. Crosses on the lines mark
on-shell intermediate particles.
\vspace {3mm}
\hspace{3 mm}
\newline
Fig.3.
Equation of the Bethe-Salpeter type for $A(t, \beta )$.
\vspace {3mm}
\hspace{3 mm}
\newline
Fig.4.
The amplitude M ( $\mu $ denotes the infrared cut-off).
\vspace {3mm}
\hspace{3 mm}
\newline
Fig.5
The evolution equation for $M$ and $\tilde M$.
\newpage
\begin{figure}
\begin{center}
\epsfig{file=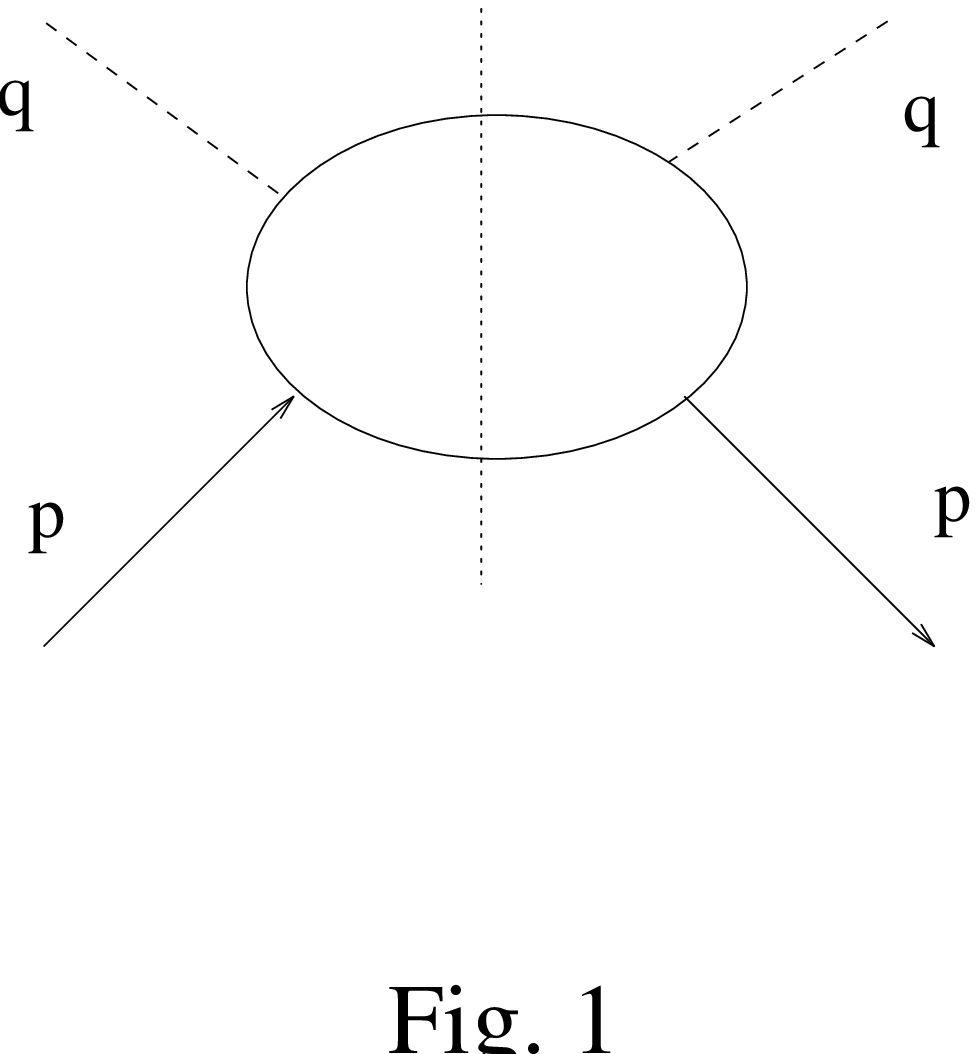}
\end{center}
\end{figure}
\begin{figure}
\begin{center}
\epsfig{file=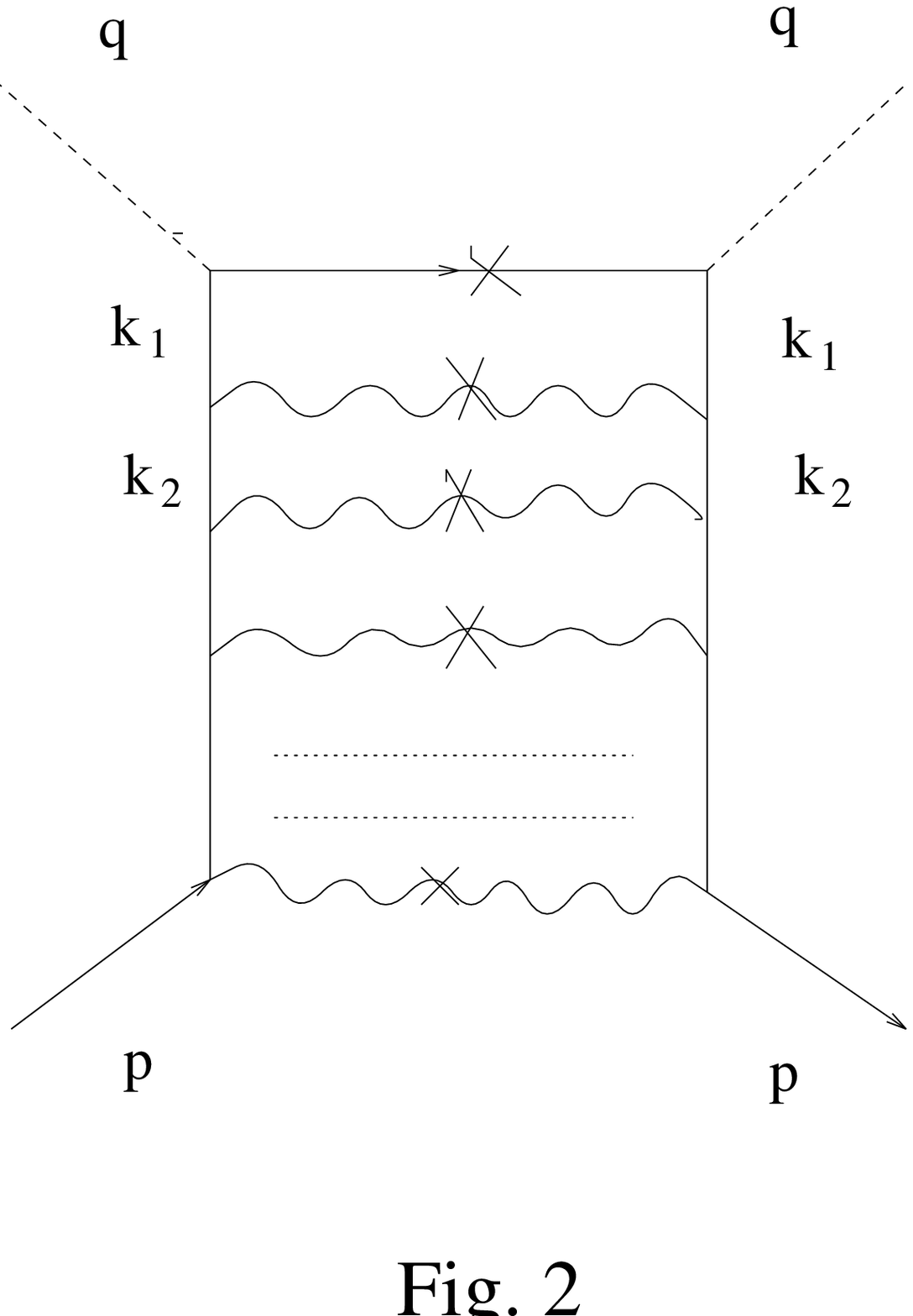}
\end{center}
\end{figure}
\begin{figure}
\begin{center}
\epsfig{file=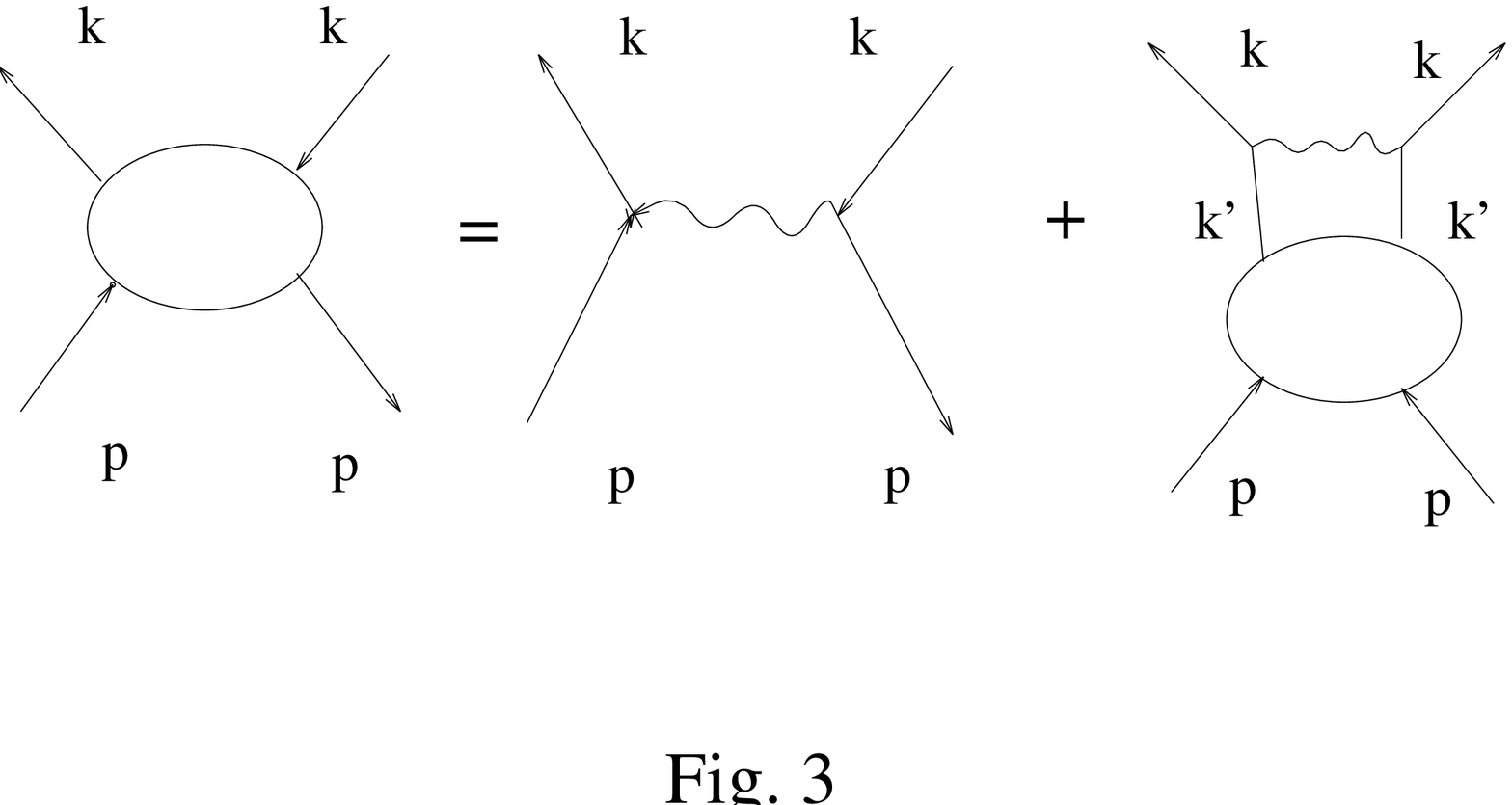,angle=270}
\end{center}
\end{figure}
\begin{figure}
\begin{center}
\epsfig{file=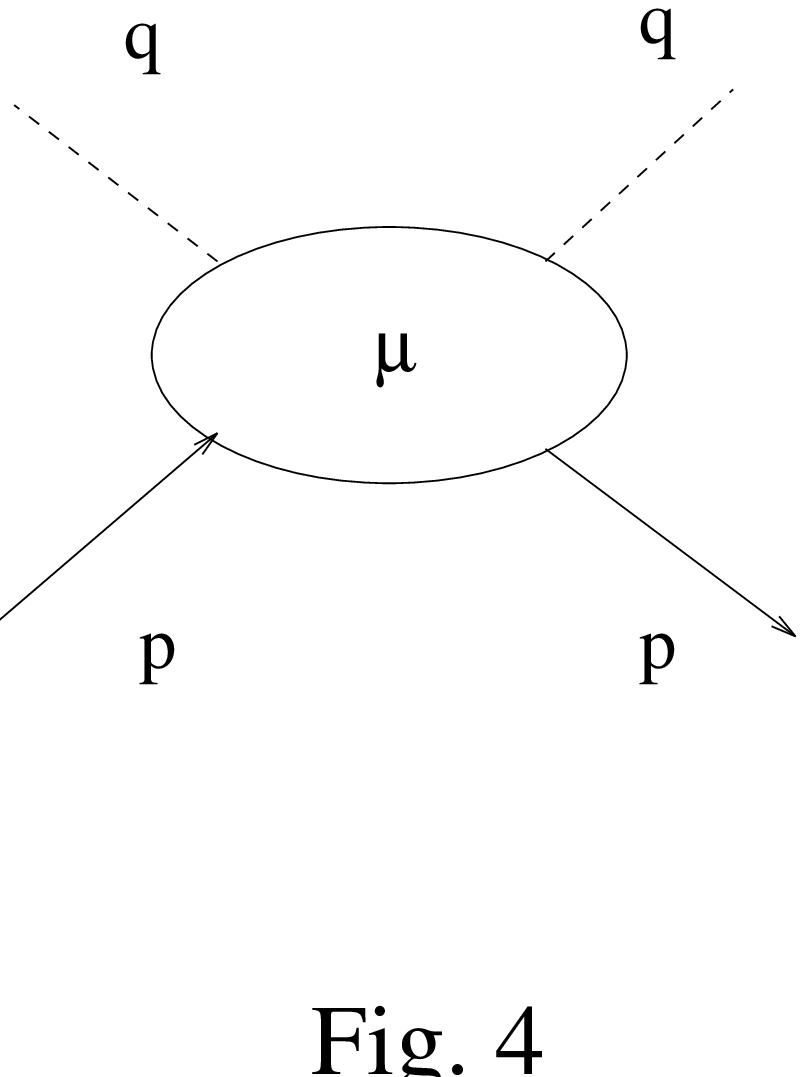}
\end{center}
\end{figure}
A
\begin{figure}
\begin{center}
\epsfig{file=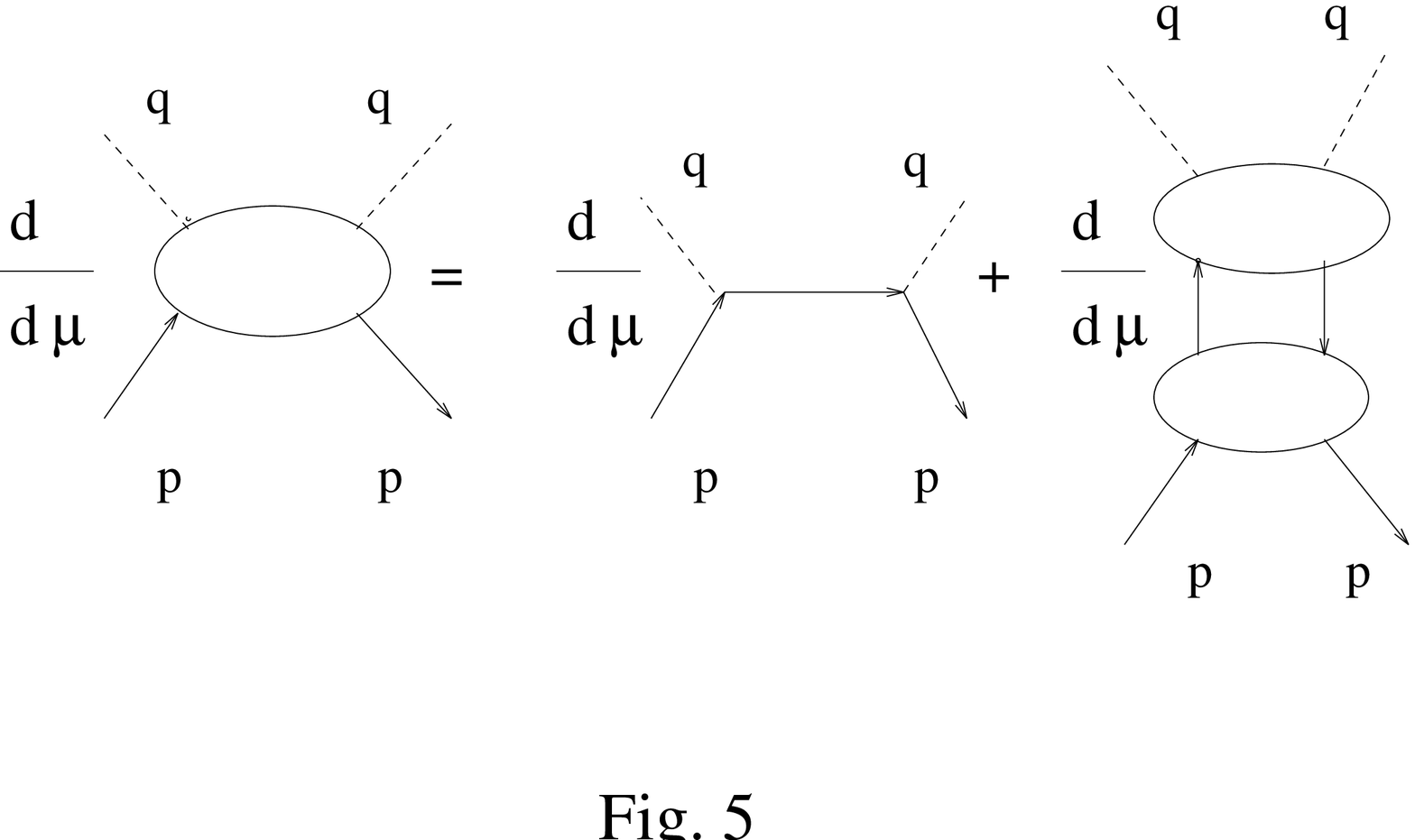,angle=270}
\end{center}
\end{figure}
\end{document}